\documentclass[12pt,preprint]{aastex}

\shorttitle{Mrk 231} 
\shortauthors{Turner, Kraemer}

\begin{document}

\title{{\it XMM-Newton} Observation of 
Fe K$\alpha$ emission from a BAL QSO: Mrk 231}

\author{T.\ J.\ Turner\altaffilmark{1,2} \&  S.\ B.\ Kraemer\altaffilmark{3}
		}
		
\altaffiltext{1}{Joint Center for Astrophysics, Physics
Dept., University of Maryland
        Baltimore County, 1000 Hilltop Circle, Baltimore, MD 21250} 
\altaffiltext{2} {Laboratory for High Energy Astrophysics, Code 662, 
	        NASA/GSFC, Greenbelt, MD 20771}
\altaffiltext{3}{Catholic University of America, NASA/GSFC, Code 681,
	Greenbelt, MD 20771}

\begin{abstract}

We present results from a 20 ksec
{\it XMM-Newton} observation of Mrk 231.
EPIC spectral data reveal strong line emission due to
 Fe K$\alpha$, which has rarely been detected in this class,  
 as BAL QSOs are very faint in
the X-ray band.  The 
line energy is consistent with an origin in neutral Fe. 
The width of the line is equivalent to a velocity dispersion 
$\sim 18,000$ km s$^{-1}$ and thus the line may be attributed to  
 transmission and/or reflection from 
a distribution of emitting clouds. 
If, instead,  the line originates in the accretion disk
then the line strength and flat X-ray 
continuum support some contribution from 
a reflected component, although 
the data disfavor a model where the hard X-ray 
band is purely reflected X-rays from a disk. 
 The line parameters are 
similar to those obtained for the Fe K$\alpha$ 
line detected in another BAL QSO, H$1413+117$.

         \end{abstract}

	         \keywords{galaxies: active -- galaxies: individual
(Mrk~231)  
       -- galaxies: nuclei -- galaxies: Seyfert} 
       
\section{Introduction - the BAL QSO phenomenon}
The ejection of matter at moderate to high velocities appears to 
be a common phenomenon in Active Galactic Nuclei (AGN). 
 Details of the accretion and ejection processes provide 
fundamental insight into the fueling of the AGN, 
 and the effect of an active nucleus on its environment. 
Moderate velocity (hundreds of km s$^{-1}$) 
outflows have been observed in many Seyfert 1 galaxies 
(e.g. \citealt{kaspi}). 
Broad Absorption Line Quasars (BAL QSOs) are 
the high-velocity systems, showing blueshifted optical and UV 
absorption lines from 
resonant transitions of ionized species such as C{\sc iv}, S{\sc iv}, 
N{\sc v} and O{\sc vi}.  \Citet{fab99} suggested that BAL 
QSOs are in an early evolutionary phase 
when the mass of gas around a black hole is beginning to get blown away 
to reveal a QSO. 

 The BAL absorption lines are thought to arise in 
radiatively-driven material along the line-of-sight, 
 with outflow velocities in the range 
5000-30,000 km s$^{-1}$ (e.g. \citealt{turn88}). BAL QSOs 
 comprise $\sim 8\%$ of the high-luminosity 
QSOs (see \citealt{wey97} for an overview). 
Models for BAL QSOs seek to explain how gas gets accelerated 
to the observed velocities. Suggestions have included a scenario where 
gas and radiation pressure lift material from the photosphere of the 
accretion disk and 
this gas is illuminated by radiation from the inner disk/corona 
\citep{m95}. That nuclear 
photon flux then accelerates the wind outwards. As the inner edge of the wind 
is thought to be quite close to the central source, some ``shielding'' 
gas has been invoked to prevent the wind material becoming stripped, and 
difficult to drive.  This gas is thought to be highly-ionized and have 
column density in the range $N_H = 10^{22} - 10^{24} {\rm cm^{-2}}$. 
\citet{prog}  found that a layer of 
shielding gas  arises naturally in their models for BAL winds. 
An alternative suggestion is that the outflow 
consists of dense clouds that are magnetically confined and that
have  small filling factor \citep{arav94}.

Constraints to date on the BAL gas location and mass-loss rates have
been based  on the UV and optical data.  X-rays may be the key to
understanding BAL QSOs as they offer the opportunity
to probe closer to the nucleus than
studies in the optical and UV bands.
Interesting X-ray results
have recently emerged with detection of  some broad absorption features 
from the lensed QSO,
APM 082769$+5255$ \citep{chartas02}, indicating an origin in gas components
outflowing at $\sim 0.2 - 0.4$c. However, BAL QSOs are very faint,
and detection of significant spectral features in the X-ray band
has been rare to date. For example, while 
Fe K$\alpha$ emission is  commonly seen in Seyfert 1 galaxies
(e.g. \citealt{n97}) and valued as an indicator of conditions
very close to the black hole, this line has been detected in 
just a few BAL QSOs, e.g.  the gravitationally-lensed 
`Cloverleaf' QSO, H$1413+117$ \citep{oshima}. 
Here we report on a brief {\it XMM-Newton} (hereafter {\it XMM})
 observation of the BAL QSO Mrk~231, yielding a rare detection of  
an Fe K$\alpha$ emission line.

\section{The Properties of Mrk 231}

Mrk~231 (z=0.042) is an ultraluminous infrared galaxy (ULIRG)  
and one of the strongest known Fe{\sc ii} emitters. 
Furthermore, Mrk~231 has one of the highest bolometric luminosities 
in the local 
(z $<$ 0.1) universe  with 
$L_{8-1000 \mu m} \sim 4 \times 10^{12} L_{\odot}$   
and $L_{BOL}> 10^{46} {\rm erg\ s}^{-1}$ \citep{soif}.  
\citet{bryantscov} find that starburst activity can 
only account for 40\% of the 
IR luminosity in Mrk 231,  an active nucleus appears to provide the 
rest of the energy. Another extreme property of Mrk~231 is its  
high polarization, $\sim 20\%$ at 2800 \AA\ (e.g. \citealt{smithea95}), 
most likely due to scattering by dust. 
Broad optical emission lines are observed similar to those 
defining the Seyfert~1 class. However, 
no narrow emission lines 
are seen except for O\verb+[+{\sc ii}\verb+]+ $\lambda 3727$. 
The absence of narrow emission lines is sometimes observed in 
quasi-stellar objects (QSOs)  
but rarely in Seyfert galaxies, this, and the high bolometric luminosity 
of Mrk~231 supported the idea that the nucleus of this source is 
similar to a QSO. 
This O\verb+[+{\sc ii}\verb+]+ emission line is consistent with 
production in gas which is 
ionized by hot young stars \citep{smithea95}  
rather than by ultraviolet radiation from the nucleus. This possibility
seems attractive  since 
the nuclear UV,  extreme-UV and soft X-ray 
 photons are likely to be heavily attenuated  
by the dense circumnuclear absorber, i.e. 
 the narrow-line gas may not 
receive ionizing radiation.

\citet{borosonea91}, 
\citet{rudyea85} and \citet{rupkeea02} are among those
discussing several distinguishable Broad Absorption Line (BAL) 
systems evident in optical and UV observations of Mrk~231. The most
prominent is defined by Na{\sc i}, Ca{\sc ii} and He{\sc i} lines
originating in gas with an outflow velocity $\sim 4200\ {\rm km\ s}^{-1}$
relative to the line emission seen in the source.
Other systems appear and disappear between observing epochs, 
 showing velocities from
$\sim 200\ {\rm km\ s}^{-1}$ inflow to 
 $\sim 8000\ {\rm km\ s}^{-1}$ outflow 
 (e.g. \citealt{borosonea91, kdh92, forstea95}).
\citet{rudyea85} showed that the shapes of 
absorption features and velocities in the 
absorbing gas systems of Mrk~231 are the same as those 
observed in BAL QSOs and 
suggested the gas is probably accelerated by the same mechanism. 
Nevertheless, the nature of the absorbing systems in Mrk~231 continued to be 
debated. 
The width of the C{\sc iv} resonance line is used as a key indicator of 
the BAL nature of an AGN \citep{wey97}.  Heavy reddening has  made 
that line width difficult to constrain in Mrk~231 until an 
{\it HST} spectrum settled 
the case in favor of a  BAL QSO designation (\citealt{g02};  hereafter G02). 
 Most  BAL systems are dominated by 
high-ionization species,  only $\sim 15$\% have absorption by 
 low-ionization systems such as Mg{\sc ii}$\lambda \lambda 2796$ and 2803, 
Al{\sc ii}$\lambda \lambda 1671$,
Al{\sc iii}$\lambda \lambda 1855$ and 1863 and 
C{\sc ii}$\lambda \lambda 1335$ \citep{vwk93}. In addition to these 
low-ionization systems, Mrk~231 shows even rarer absorption lines, such 
as Na{\sc i}D, He{\sc i}$\lambda3889$ and Mg{\sc i}$\lambda 2853$.  
While Mrk~231 does show some of the high-ionization systems (G02) it 
is dominated by the low-ionization systems; 
 the Mg{\sc ii} absorption in particular, places Mrk~231 into the 
so-called LoBAL subclass of low-ionization BAL QSOs.

\subsection{Previous X-ray Observations of Mrk 231}

The {\it ROSAT} High Resolution Instrument (HRI)
showed extended soft X-ray  emission
consistent with the extent of the host galaxy ($\sim 10''$; \citealt{t99}),
confirmed using {\it Chandra} (G02).
An early {\it ASCA} observation \citep{i99,t99}
showed a flat spectrum but with insufficient
counts to distinguish between  absorption or
reflection models.
In this paper we use the term ``reflection'' to refer to 
reprocessing of photons 
via  Compton scattering and fluorescence by material 
which is optically-thick to electron scattering. We use the term 
``scattering'' to refer to the case where the optical depth to 
electrons is $0< \tau < 1$.  
A longer {\it ASCA} observation
(\citealt{mr}; hereafter MR) suggested the X-ray spectrum to
be dominated by reflected plus scattered nuclear emission
with a large absorbing column in the line-of-sight.
Thus MR suggested a 
Seyfert 2 type geometry and orientation for Mrk~231,  with
the accretion-disk system  surrounded by a 
shell of absorbing gas $N_H \sim 3 \times 10^{22} {\rm cm^{-2}}$ 
which covers direct and reflected/scattered components. The shell  
may  block ionizing radiation from the 
optical narrow-line region, while allowing hard X-rays to penetrate. 
G02 find significant variability in the hard X-ray flux,
 and suggested
nuclear  X-rays are hidden by a Compton-thick cloud (Fig~12 of G02) existing
on size-scales $< 10^{15}$cm, while 
scattered X-rays from multiple lines-of-sight reach the observer. 
In the context of the MR model the hard X-rays could vary if the 
reprocessor were suitably small. This  may be reasonable for  a source 
like Mrk 231,   
radiating at a large fraction of the Eddington luminosity, as the inner 
disk may puff up into a funnel \citep{rees84} yielding a reflector 
surface consistent with the size 
implied by variability in this case.   
\citet{smithea95} suggest there may be 
a superposition of two polarized components in Mrk~231, supporting the 
importance of scattering in this system. 

\section{The XMM-Newton Observation}
\subsection{Data Reduction}

An {\it XMM} observation  of Mrk~231 
was performed starting 2001 June 7 UT 13:10:24 
with an exposure time of $\sim 22$ ks.  
Here we present analysis of the archived {\it XMM} data 
from that observation.  {\it EPIC} data were processed  
 using  SASS 5.4. EPIC pn and MOS data utilized the medium filter and 
the prime full window. The EPIC data were screened to remove hot and bad 
pixels, and some brief periods of high background counts. Instrument patterns 
0 -- 12 (MOS) and 0 -- 4 (pn) were selected. 
The screening criteria resulted in an effective exposure of 
21.4 ks for the MOS and pn data. 
  The  background-subtracted count rates were
  $0.0391\pm0.0014$ (MOS1), $0.0389\pm0.0014$ (MOS2)  
 and $0.130\pm0.0028$ (pn) over the 0.3-10.0 keV band.  
Spectra and time series were extracted from a cell of radius 
$\sim 30''$ centered on the source. Background spectra were extracted 
from nearby, source-free regions of MOS and pn, the background was 
$\sim 6$\% of the source count rate.   
Unfortunately this brief exposure accumulated only
400 source counts in the     
combined arms of the RGS data,  insufficient 
for spectral analysis.

\subsection{EPIC Results}

{\it XMM} data showed Mrk~231 to have an observed  flux
$F_{2-10\ keV} \sim 6.2 \times 10^{-13} {\rm erg\ cm^{-2}\ s^{-1}}$, 
during the June 2001 observation, corresponding to an  
observed luminosity 
$L_{2-10\ keV} \sim 2 \times 10^{42} {\rm erg\ s^{-1}}$ 
(assuming $H_0=75$ km s$^{-1}$ Mpc$^{-1}$,  $q_0=0.5$ throughout). 
The 2 - 8 keV flux 
($5 \times 10^{-13} {\rm erg\ cm^{-2}\ s^{-1}}$) 
is consistent with that reported by G02.  
We found no significant variations in X-ray flux within this short 
 {\it XMM} exposure. 

\begin{figure}[h]
\epsscale{0.65}
\plottwo{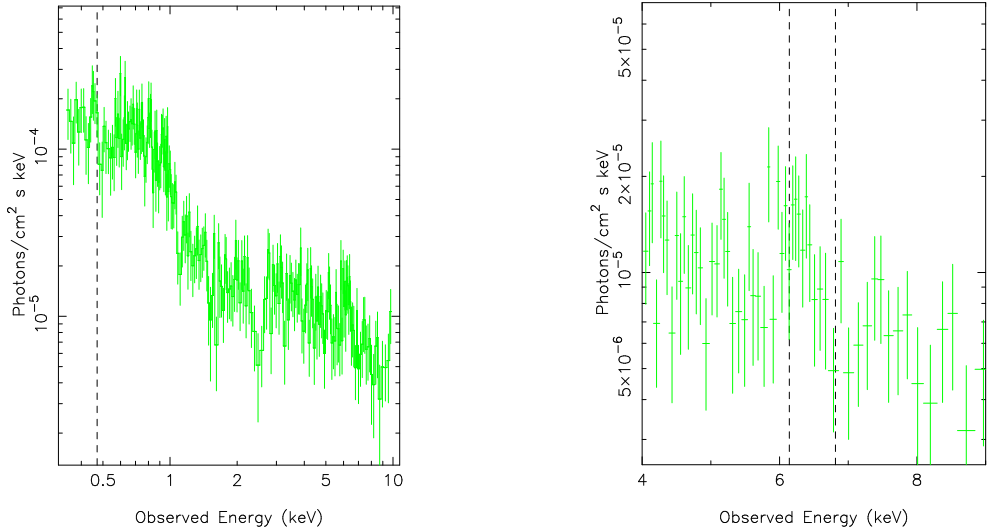}{f1b.eps}
\caption{\it Left: The EPIC pn spectrum of Mrk~231 from a 20 
ksec exposure with XMM. 
A large flux of lines is also evident between 
  0.5 - 1 keV. A sharp dip just above 2 kev 
does not show up in MOS data and is likely to be an artifact of the 
instrument.  The dashed vertical line shows where we would expect 
the C{\sc vi} RRC feature, at the redshift of the host galaxy. The right 
panel shows a close-up of the Fe K$\alpha$ line. Dashed lines denote  
the expected placement of a line emitted at 6.4 keV and an 
edge at 7.1 keV, observed at the redshift of the host galaxy.  }  
\end{figure}

The overall spectral shape is complex. 
Several previous papers have discussed the difficulty in distinguishing 
between  models where the active nucleus is 
covered by a patchy and/or ionized  
absorber \citep{i99,t99} and those 
where the spectrum is dominated by a reflected component (MR). This 
{\it XMM} exposure is brief and 
the data do not allow us to determine the precise contributions 
of reflected and absorbed continua to the overall spectrum.  
 However, progress is possible as the pn spectrum shows two prominent
 emission lines (Fig~1) plus emission consistent with a blend of lines  
from the (previously suspected) thermal component peaking between 0.5 - 1 keV 
(\citealt{i99,t99,g02}, MR).
EPIC data  show some evidence for
absorption-like features at $\sim 1.1$, 1.6  and
$\sim 2.4$ keV. The latter appears only in 
pn data and is probably an instrumental effect. For the others, 
 the low S/N combined with 
 the modest energy resolution leads to  
 ambiguity between emission and absorption features.  
 These may be absorption features due to Fe,  
 but these data do not allow an unambiguous determination. 

One of the most obvious features in the pn data 
is immediately identified with emission from the K-shell of Fe. 
The pn detects 30 counts in the line during this exposure. 
The MOS spectra have far fewer counts and the line is not significantly
detected in those (although the data are consistent with the pn 
results). Thus we used only the pn data to derive the line parameters 
given below. Detection of an Fe line in the EPIC spectrum from Mrk~231 is 
also reported within the analysis of a sample of 
Ultra-Luminous Infrared Galaxies performed by \citet{fran}. 

\begin{figure}[h]
\epsscale{0.65}
\plottwo{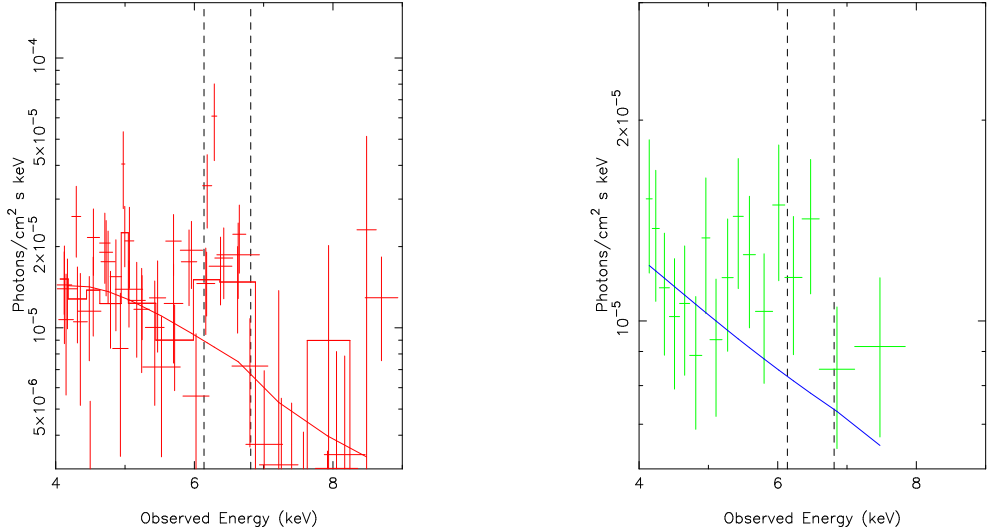}{f2b.eps}
\caption{\it  The left-hand panel 
 the Fe K$\alpha$ line in the combined instruments provided by ASCA data, 
the dashed vertical lines represent  the energies 
of a line at 6.4 keV and an edge at 7.1 keV 
observed at the redshift of the host galaxy. 
The right-hand panel shows data in the same regime from our analysis of the 
archived ACIS observation of 2000 October. 
The Fe K$\alpha$ line is evident in both datasets, and the flux is consistent 
with that found by the pn. }  
\end{figure}

First we fit the Fe K$\alpha$ line relative to a `local' continuum 
parameterized as a powerlaw between 4 -- 9  keV,
 absorbed by the galactic line-of-sight column 
($N_H= 1.27 \times 10^{20}\ {\rm cm^{-2}}$). 
This yielded $\Gamma \sim 1.01^{+0.39}_{-0.38}$, indicative 
of complex absorption and/or a contribution from reflection, as 
previously found for this source. We measured an observed-energy 
for the emission line 
$E=6.23\pm0.12$ keV,  corresponding to a rest-energy $E=6.49\pm0.12$ keV  
correcting for the systemic velocity of the host galaxy.
The line width was $\sigma=0.17^{+0.12}_{-0.11}$ keV with flux 
$n=3.85^{+2.11}_{-1.86} \times 10^{-6}$ 
photons cm$^{-2}$ s$^{-1}$ and the luminosity in the line was 
$L_{FeK}=1.3 \times 10^{41}{\rm erg\ s^{-1}}$.  The lower limit of 
$\sigma=60$ eV means the Fe K$\alpha$ line has a 
significant breadth.   (Errors are 90\% confidence.) 
The line equivalent width is $\sim 450^{+229}_{-204}$  
eV relative to the local  continuum.  
For comparison the spectral resolution of the pn at 6.4 keV is 
$\sigma \sim 50$ eV. Fig~1 shows the data compared to lines denoting 
the expected observed energies of an emission line from neutral Fe, 
along with the energy of the Fe K edge which would 
arise in the  same gas, both assumed to be observed at the redshift 
of the host galaxy. While a dip is evident at the 
energy where an edge is expected, 
addition of an edge did not improve the fit significantly. 

For comparison, we fit the line relative to a ``pure 
reflection'' model for data in the 2 - 10 keV band (as suggested by MR). 
This fit utilized the {\sc pexrav} model  \citep{mzdz95} from 
{\sc xspec}. Fits were attempted assuming disk inclination 
angles of 45$^{\rm o}$ and $0^{\rm o}$ (face-on). The (unseen) powerlaw 
was assumed to be $\Gamma=2.0$ and to turn-over 
at 100 keV. 
Fits using the former disk orientation 
gave $\chi^2=103/97\ dof$ and 
a line equivalent width $EW=202^{+208}_{-202}$ eV 
relative to the reflected continuum, while 
a face-on disk yielded $\chi^2=103/97\ dof$, 
$EW=139^{+221}_{-139}$. 
The best-fit equivalent width in either case is lower than 
that found relative to the absorbed powerlaw because the
 edge in the reflected continuum contributes to the curvature 
just above 6.4 keV, and so fewer photons 
are attributed to the line. In these two (pure reflection) disk models 
 the best-fit line strength is about 10\% of the 
strength expected, relative to the reflected continuum \citep{fgf}. 
Thus the line strength is low compared to the model in this picture. As 
the discrepancy is in the ratio of reflected line to reflected 
continuum, we 
can extend this result to  a range of geometries. 
The line is too weak relative to the reflected continuum 
if reflection occurs from either a disk or a distribution of clouds. 

Alternative fits were attempted, which utilized both 
 transmitted and reflected components, with an unconstrained absorber 
covering both. First we considered the face-on disk. 
 We fixed the reflection to that expected from a flat 
disk illuminated by an isotropic source ($R=1$ in {\sc pexrav}). This 
yielded $\chi^2=93/97\ dof$ with 
a flat photon index $\Gamma=1.14^{+0.48}_{-0.75}$,  
$N_H=2.69^{+2.41}_{-1.57} 
 \times 10^{22}\ {\rm cm^{-2}}$, indicating 
additional, but undeterminable complexity to the absorber. 
We obtained similar results for a disk at 45$^{\rm o}$, with 
 $\chi^2=93/97\ dof$, $\Gamma=1.14^{+0.47}_{-0.74}$, 
$N_H=2.71^{+2.40}_{-1.96}  \times 10^{22}\ {\rm cm^{-2}}$.  
Fixing, instead, the 
photon index at $\Gamma=2$ yielded $R \sim 10$ for both disk orientations, 
equivalent to the 
`pure reflection' solutions, above.

To confirm the presence of the Fe line, and look for flux variability we 
reduced the long archival observation by {\it ASCA}, from 1999 November 
10-12. Data reduction followed 
the methods outlined for the Tartarus database \citep{tart}. The 
{\it ASCA} spectral data were fit using all four instruments, 
and the spectrum confirms the presence of the Fe line (Fig~2). 
The spectrum was modeled 
constraining photon index, absorption, line energy and line width to match 
the values derived from the {\it XMM} data. This produced a good fit, with 
continuum flux 
$F_{2-10\ keV} \sim 7.5 \times 10^{-13} {\rm erg\ cm^{-2}\ s^{-1}}$ and 
line flux $n=5.76^{+2.91}_{-2.78} \times 10^{-6}$ 
photons cm$^{-2}$ s$^{-1}$. 

We also reduced the archived ACIS data from 2000 October 19-20, 
originally reported by G02 (Fig~2). 
 Extracting ACIS CCD spectra of Mrk~231 
we fit the data using the same continuum and line model used for the 
{\it XMM} and {\it ASCA} spectra, 
leaving only line and continuum normalizations free. This model
 yielded a line flux 
 $n=2.48^{+2.26}_{-2.25} \times 10^{-6}$ 
photons cm$^{-2}$ s$^{-1}$. The differences in line flux are not significant 
and so the data 
show the line flux to be consistent with a constant value 
over a baseline of 1.5 years.

Turning to the soft X-ray band. 
Previous PSPC \citep{t99} and {\it Chandra} ACIS 
data (G02)  allowed limited spatially-resolved spectroscopy which 
provided two-temperature thermal models 
for the extended soft emission. While previous models account for the 
approximate flux in the soft band, they fail to produce some observed 
features such as the relatively isolated peak evident at 
$0.471\pm0.03$ keV (Fig~1). 
Correcting for the redshift of the host galaxy we obtain a
rest-energy 0.490 keV, nominally 
consistent with C{\sc vi} RRC 0.492 keV, although most of the line flux lies 
slightly redward of this energy (Fig~1). 
 Confirmation 
of C{\sc vi} RRC would be interesting as the width 
would constrain temperature 
and excitation mechanism for the gas. 

\section{Discussion} 

A brief {\it XMM} observation of Mrk~231 has revealed 
a rare 
 detection of Fe K$\alpha$ from a BAL QSO. The presence of 
strong Fe K$\alpha$ emission dispels any remaining doubt as to the 
presence of an active nucleus at the heart of Mrk~231. 
The existence of this 
line is confirmed by analysis of archived {\it ASCA} and {\it ACIS} data, 
and 
the line flux is consistent with a constant value between all  
datasets, covering a baseline of $\sim 1.5$ years. 
The luminosity in the line is $L_{Fe} \sim 10^{41} {\rm erg\ s^{-1}}$ 
which is typical of the Fe K$\alpha$ luminosities observed in Seyfert 
1 and 2 galaxies (e.g. \citealt{n97,t97}). Assuming a 
nuclear system with an intrinsic  luminosity in the Seyfert 1 range, the 
luminosity of the line would disfavor any model where we are seeing the 
line via scattering from optically-thin gas, which typically 
scatters only a few percent of nuclear emission into our line-of-sight 
in Seyfert 2 galaxies.  The other alternative is  that  
we see only a few percent of the line flux, which happens to come out 
to a line luminosity typical of Seyfert 1 galaxies. In this case the 
nuclear luminosity would just have to be a factor 50 or so higher than 
a Seyfert nucleus. Different analyses have yielded a wide range 
of inferred nuclear luminosities for  Mrk~231, 
including an estimate at $2 \times 10^{44} {\rm erg\ s^{-1}}$ (MR) 
which encompasses the latter possibility.
The high equivalent width of the line, and the flat continuum shape 
are reminiscent of the properties of H$1413+117$ \citep{oshima}. 
A line of this equivalent width could be produced 
from shell of gas 
$N_H \sim 3 \times 10^{23} {\rm cm^{-2}}$ \citep{lc93}. This column 
is consistent with  that found when the X-ray spectrum is fit with a 
powerlaw, partially-covered by cool gas \citep{t99}. The line is too 
strong to be produced by scattering from a  shell  
with $N_H \sim few \times 10^{22} {\rm cm^{-2}}$, as suggested to cover  
the central regions (MR).

We turn now to the line energy. 
We consider the possibility that the line arises from neutral Fe.  If so,
    there is no systematic velocity shift for the emitting gas. While the 
data  disfavor models where we preferentially observe one side 
of a flow system, the line breadth 
is equivalent to a velocity-width 
of $18500^{+13000}_{-12000}$ km s$^{-1}$. The 
geometry of the emitting gas remains unconstrained, it could be 
anything from a   
spherical distribution  or ring of clouds to the conical 
BAL flow, so long as we see contributions from both sides   
(as the line peak is not shifted). 
A specific model which fits this general picture is that  
 of \citet{arav94}, where the line would arise in 
 dense clouds with a small filling factor. The   
Fe K$\alpha$ emission in H$1413+117$ \citep{oshima,g02b}   
 was also found to be consistent with an origin in neutral gas. 
In that case, emission from the BAL flow was used to explain the line,  
with the width suggesting a flow velocity of 12,000 km s$^{-1}$. 

   An alternative is that 
    the line is produced in the accretion disk and that 
 the broadening can be attributed to
  strong gravity, as it is in the spectra of Seyfert type AGN, where 
there are no high-velocity BAL flows. Of course the line width  observed here 
is rather modest compared to the distortion 
observed in Seyfert profiles.  
The equivalent width of the line is $\sim 450$ eV, 
about twice that expected
from a source isotropically illuminating a flat disk \citep{fgf}, assuming 
we see both the primary and the reflected components.    This may 
   infer the reprocessor sees more of the illuminating flux than
     the observer does, or that there is a lag between primary and
     reprocessed spectra.  An alternative explanation of the line strength 
is that the Fe abundance is very
 high in BAL QSOs, as expected if these are young systems \citep{hamfer}.

However, fits  assuming we see only  the 
reflected spectrum in the 2 - 10 keV band 
(i.e. the  primary continuum is hidden) show  
the Fe line has just 10\% of the strength expected, 
relative to the reflected continuum. This  
 supports the case for some  contribution from the primary continuum 
in the hard X-ray regime and thus disfavors pure reflection models, 
whether reflection occurs from a disk or from a distribution of clouds. 
 
Of course, the breadth of the line could be due simply to Compton 
scattering of line photons in material along the line-of-sight,  
as  discussed during the early days of Fe K$\alpha$ observations 
from Seyfert 1 galaxies (e.g. \citealt{mz95}). If so, 
 the lack of shift to the peak indicates the opacity of
that gas is low, while the upper limit on line width, 
$\sigma$,  constrains the gas to $kT < 0.5 $ keV 
(using $\Delta E/E= \sqrt{2kT/mc^2}$), following \citealt{mz95}). 

 Finally, it is impossible to rule out 
emission from a higher energy, redshifted down to 6.49 keV.  
A velocity of 24,000 km s$^{-1}$ is possible if the line is from
 recombination in H-like Fe. However, if the gas were highly ionized 
we would most likely see emission from both 
He-like and H-like Fe (and the observed feature would have to be a blend of 
these two). Further, it seems unlikely 
H$1413+117$ and Mrk~231 would both show single broad 
lines consistent with 6.4 keV by this conspiracy. Thus we favor 
an origin in neutral gas.

\section{Acknowledgements}

We are grateful to the  {\it XMM}   
guest observer support team who provided useful input 
regarding the analysis of these data. We thank Ian George for 
useful comments.  T.J.\ Turner acknowledges support from NASA 
grant  NAG5-7538.


\begin{thebibliography}{}       

\bibitem[Arav, Li \& Begelman(1994)]{arav94}
	Arav, N., Li, Z. Y., Begelman, M. C., 1994, \apj, 432, 62 
\bibitem[Boroson et al.(1991)]{borosonea91}
	Boroson, T.A., Meyers, K. A., Morris, S. L., Persson, S. E., \
	 1991, \apjl, 370, L19 
\bibitem[Bryant \& Scoville(1996)]{bryantscov}
	Bryant, P.M., Scoville, N.Z. 1996, \apj, 457, 678 
\bibitem[Chartas et al.(2002)]{chartas02}
	Chartas, G., Brandt, W. N., Gallagher, S. C., 
	Garmire, G. P.,  2002, \apj, 579, 169  
\bibitem[Fabian(1999)]{fab99}
	Fabian, A. C., 1999, \mnras, 308, 39
\bibitem[Forster et al.(1995)]{forstea95}
	Forster, K., Rich, R. M., McCarthy, J. K., 1995, \apj, 450, 74 
\bibitem[Franceschini et al.(2003)]{fran}
	Franceschini, A., et al., 2003, astro-ph/0304529 
\bibitem[Gallagher et al.(2002)]{g02}
	Gallagher, S.C., Brandt, W.N., Chartas, G., Garmire, G. P., 
	Sambruna, R. M.,  2002 (G02)  \apj 569, 655 
\bibitem[Gallagher et al.(2002b)]{g02b}
	Gallagher, S.C., Brandt, W.N., Chartas, G., Garmire, G. P.
	 2002b,  \apj 567, 37
\bibitem[George \& Fabian(1991)]{fgf}
	George, I. M. \& Fabian, A. C., 1991, \mnras, 249, 351 
\bibitem[Hamann \& Ferland(1999)]{hamfer}
	Hamann, F, Ferland, G. 1999, ARA\&A 37, 487 
\bibitem[Iwasawa(1999)]{i99}
	Iwasawa, K. 1999, MNRAS 302, 96 
\bibitem[Kaspi et al.(2002)]{kaspi}
	Kaspi, S. et al. 2001, \apj, 574, 643 
\bibitem[Kollatschny, Dietrich \& Hagan (1992)]{kdh92}
	Kollatschny, W., Dietrich, M., Hagan, H., 1992, \apjl, 264, L5 
\bibitem[Leahy \& Creighton(1993)]{lc93}
	Leahy, D. A., Creighton, J., 1993, \mnras, 263, 314 
\bibitem[Magdziarz \& Zdziarski(1995)]{mzdz95}
	Magdziarz, P., Zdziarski, A. A., 1995, \mnras, 273, 837
\bibitem[Maloney \& Reynolds(2000)]{mr}
	Maloney, P. R., Reynolds, C. S., 2000 (MR) ApJ 545, 23 
\bibitem[Murray et al.(1995)]{m95}
	Murray, N., Chiang, J., Grossman, S. A., Voit, G. M., 
	1995, \apj, 451, 498 
\bibitem[Mushotzky et al.(1995)]{mz95}
	Mushotzky, R. F., Fabian, A. C., Iwasawa, K., Kunieda, H., 
	Matsuoka, M., Nandra, K., Tanaka, Y., 1995, \mnras, 272, 9 
\bibitem[Nandra et al.(1997)]{n97}
	Nandra, K., George, I. M., Mushotzky, R. F., Turner, T. J., 
	Yaqoob, T., 1997, \apj, 477, 602 
\bibitem[Oshima et al.(2001)]{oshima}
	Oshima, T., et al. 2001, \apj, 563, L103 
\bibitem[Proga et al.(2000)]{prog}
	Proga, D., Stone, J. M., Kallman, T. R., 2000, \apj, 543, 686
\bibitem[Rees(1984)]{rees84}
	Rees, M.J.,1984, ARAA, 22, 471 
\bibitem[Rudy et al.(1985)]{rudyea85}	
	Rudy, R.J., Stocke, J. T., Foltz, C. B.,  1985, \apj, 288, 531 
\bibitem[Rupke et al.(2002)]{rupkeea02}
	Rupke, D. S., Veilleux, S., Sanders, D. B.,  2002, \apj 570, 588 
\bibitem[Soifer et al.(1987)]{soif}
	Soifer, B.T., Sanders, D. B., Madore, B. F., Neugebauer, G., 
	Danielson, G. E., Elias, J. H., Lonsdale, C. J., Rice, R. W., 
	1987  \apj, 320, 238 
\bibitem[Smith et al.(1995)]{smithea95}	
	Smith, P.S., Schmidt, D. G., Allen, R. G., Angel, J. R. P., 
	1995, \apj, 444, 146 
\bibitem[Turner et al.(1997)]{t97}
	Turner, T.J., George, I. M., Nandra, K., Mushotzky, R. F., 
	1997, \apjs, 113, 23 
\bibitem[Turner(1999)]{t99}
	Turner,T.J. 1999,\apj\ 511, 142 
\bibitem[Turner et al.(2001)]{tart}
	Turner, T.J., Nandra, K., Turcan, D., George, I. M.,  in 
	``X-ray Astronomy: stellar endpoints, AGN and the diffuse X-ray 
	background''. Ed. N. E. White, G. Malaguti, G.G.C. Palumbo,
	2001, Melville, NY, AIP, 599, 1041.  
\bibitem[Turnshek et al.(1988)]{turn88}
	Turnshek, D. A., Grillmair, C. J., Foltz, C. B. \& Weyman, R. J., 
	1988, \apj, 325, 651  
\bibitem[Voit, Weymann \& Korista(1993)]{vwk93}
	Voit, G. M., Weymann, R. J., Korista, K. T.,  1993, \apjs, 88, 357 
\bibitem[Weyman(1997)]{wey97}
	Weyman, R.J., 1997 in ASP Conf. Ser. V.128 	        
\end{thebibliography}
\end{document}